\documentclass[amsmath]{aastex702}

\usepackage{url}

\usepackage{graphicx}
\usepackage{color}
\usepackage{siunitx}
\usepackage{hyperref}
\usepackage{longtable}
\usepackage{overpic}

\hypersetup{
	colorlinks,
	citecolor={blue},
	urlcolor={red},
	linkcolor={blue},
}

\usepackage{float}
\usepackage{amsmath}

\begin{document}
	
	\title{Energy-Dependent Hard X-Ray QPP Modulation in M- and X-Class Solar Flares Observed by Aditya-L1/HEL1OS}
	
	\author[orcid=0009-0009-3048-0109,sname='Kanaujiya']{Sachin Rajkapoor Kanaujiya}
	
	\affiliation{National Institute of Technology Calicut, Kozhikode-673601, Kerala, India}
	\email[show]{sachin\_p220261ph@nitc.ac.in}

	\author[orcid=0000-0002-6353-9252,sname='Maurya']{Ram Ajor Maurya}
	\affiliation{National Institute of Technology Calicut, Kozhikode-673601, Kerala, India}
	\email[show]{ramajor@nitc.ac.in}

	\author[orcid=0000-0001-6423-8286,sname='Nakariakov']{Valery M.	Nakariakov}
	\affiliation{Centre for Fusion, Space and Astrophysics, Department of Physics, University of Warwick, Coventry CV4 7AL, UK}
	\affiliation{G-LAMP NEXUS Institute/School of Space Research, Kyung Hee University,  Yongin, 17104, Republic of Korea}
	\affiliation{Centro de Investigacion en Astronom\'ia, Universidad Bernardo O'Higgins, Avenida Viel 1497, Santiago, Chile}
	\email[show]{V.Nakariakov@warwick.ac.uk}

	\shorttitle{QPPs in X-Class Flares}
	\shortauthors{Kanaujiya, et al. 2026c}
	
	\begin{abstract}

	Quasi-periodic pulsations (QPPs) in solar-flare hard X-ray (HXR) emission provide observational constraints on the temporal variability of flare energy release and energetic-electron production. In our previous study, we established a systematic catalogue of HXR QPPs in HEL1OS observations of X-class solar flares, identifying statistically significant pulsations in 25 of 34 events with sufficient data quality \citep{2026arXiv260922787R}. Here, we extend this investigation by comparing the energy dependence of the modulation depth of QPPs in M- and X-class flares across four HEL1OS energy channels spanning 8–20 keV. The comparison comprises 18 M-class events and the X-class sample from our previous study, with the number of available measurements varying between energy channels. In both flare populations, the modulation depth increases with photon energy. For M-class flares, it rises from approximately 2.3\% at 8–11 keV to 13.2\% at 17–20 keV, whereas for X-class flares increase from approximately 1.0\% to 3.1\%. The stronger increase towards higher energies in the M-class sample constitutes the principal observational contrast. These results demonstrate an energy-dependent increase in the relative amplitude of HXR QPPs in both populations and motivate a systematic statistical assessment of whether their energy-dependent modulation-depth distributions differ. The comparison provides a basis for constraining models of QPP generation and the modulation of energetic-electron emission during solar flares.

	\end{abstract}
	
	
	\section{Introduction}
	\label{sec-intor}

	Quasi-periodic pulsations (QPPs) that recur at nearly, but not exactly, regular intervals are a frequent feature of flares on the Sun and on other stars. These QPPs has a range from a fraction of a second to tens of minutes, and they have been recorded at wavelengths ranging from radio and optical/EUV to soft and hard X-rays and gamma-rays \citep[e.g.,][]{Parks1969, 1987ApJ...321.1031T, Asai2001,Nakariakov2003a, Foullon2005,Ofman2006,Li2008,Srivastava2008,Nakariakov2009,Nakariakov2010,Yuan2011,Yang2016,Hayes2016,Hayes2019,Kupriyanova2020,2026ApJ..1003...78K}. Pulsations are seen both in emission from heated plasma (soft X-rays, EUV) and in emission from energetic particles (hard X-rays, microwaves), which is why they are thought to carry information on how flares generate energy and accelerate electrons.
	
	The earliest evidence came from radio and X-ray records obtained in the 1960s and 1970s, in which pulses with periods of the order of ten seconds were noticed in microwave bursts \citep[e.g.,][]{Parks1969, Aschwanden1987}. Hard X-ray measurements by Yohkoh \citep[e.g.,][]{1991yohk.book.....S} later disclosed pulse trains with periods between roughly 10 and 100~s in large events \citep{2014CEAB...38..111T, Inglis2015, 2015SoPh..290..115S, 2019AdSpR..64.1100S, 2022ApJ...936...87F}. Studies of strong flares with RHESSI showed that more than one period can be present at once: \citet{Inglis2009}, analysing the X3.9 flare of 3 November 2003, found pulsations at about 12~s and 28~s and proposed that they arise from different MHD modes of the flaring structures or from oscillatory reconnection. Microwave imaging from the Nobeyama Radioheliograph produced a further set of cases, among them 15--30~s gyrosynchrotron pulsations that closely track hard X-ray bursts, which ties the phenomenon to the acceleration of non-thermal electrons \citep[e.g.,][]{Nakariakov2003a, Tan2010}.
	
	Since 2010, SDO has provided EUV images at fine spatial and temporal resolution through the Atmospheric Imaging Assembly \citep[AIA;][]{2012SoPh..275....3P, lemen2012aia}, giving a new impetus to the field. Combining several wavebands, \citet{Simoes2015} found that EUV, soft X-ray and microwave pulsations vary in step, implying one underlying driver. Pulsations lasting for a long time, with periods of 50--60~s, were found in X1.6 and X2.0 flares using AIA together with Fermi/GBM \citep{Hayes2019, Hayes2020}. In the X9.3 flare of 6 September 2017, the largest event of Cycle 24, thermal emission measured by GOES/XRS and SDO/EVE contained two simultaneous pulsation types, with periods of 12--25~s and 4--5~min \citep{2018ApJ...858L...3K}. Observations like these indicate that pulsations accompany a large share of major flares. Moreover, \citet{2016ApJ...830..110C} compared RHESSI and XMM-Newton soft X-ray records and showed that solar and far more energetic stellar flares host QPPs of a similar kind.
	
	A wide range of explanations has been proposed; thorough overviews are given by \citet{McLaughlin2018}, \citet{Kupriyanova2020} and \citet{Zimovets2021}. The spread in periods, the varied shapes of the pulsation envelope and the non-stationarity of the signals suggest that more than one process contributes. The proposals can be sorted into three categories. In the first, an MHD oscillation modulates the emitting plasma or the motion of non-thermal electrons produced in the flare \citep[e.g.,][]{2008PhyU...51.1123Z, 2012ApJ...761..134N, 2022MNRAS.516.2292K}. In the second, the electron acceleration is itself pulsed, because a wave inside or close to the flare region, or one arriving from the chromosphere, modulates the rate of magnetic reconnection \citep[e.g.,][]{Nakariakov2016, Sych2009}. In the third, reconnection is intrinsically bursty, so that heating and particle acceleration recur by themselves. These mechanisms are not mutually exclusive, and their coexistence may account for flares in which pulsations of very different periods occur together \citep[e.g.,][]{Kolotkov2015, 2017MNRAS.471L...6L, 2018ApJ...858L...3K, 2021ApJ...921..179L, 2022SoPh..297....2N, 2025JGRA..13033772L}.
	
	Among the measurable quantities describing a QPP, the modulation depth, defined as the amplitude of the pulsation relative to the flare emission \citep[e.g.,][]{Nakariakov2003a, Nakariakov2010, Li2024}, is especially informative. It tells us how strongly the driver modulates the emitting plasma or the accelerated electrons, and it helps to establish whether the modulation is linear or nonlinear. Our previous work \citep{2026arXiv260922787R} indicated that pulsations tend to be stronger at higher photon energies, or in non-thermal rather than thermal emission \citep{Li2008, NakariakovMelnikov2009, Kupriyanova2010, Hayes2016, Li2017}. Some reviews state that this is what one would expect if energetic electrons are injected in pulses by repeating reconnection \citep{Kupriyanova2020, McLaughlin2018}. However, modulation depths are mostly reported for single events. How they vary with photon energy, flare magnitude, pulsation period and pulsation duration has not been constrained well by statistical work, and large samples have generally been used to ask how often QPPs occur rather than how strong they are \citep[e.g.,][]{2015SoPh..290.3625S, 2019AdSpR..64.1100S, 2020ApJ...895...50H}.
	
	Accurate measurements of the depth also depend on how the data are treated. Flare light curves are strongly non-stationary, and an amplitude averaged over the whole flare rather than over the stretch where oscillations are truly present will be biased. Wavelet transforms, the usual detection tool, work best for signals that are locally harmonic, yet QPPs are often non-harmonic \citep[e.g.,][]{Nakariakov2019}. Few authors limit their amplitude estimates to time intervals of statistically significant wavelet power, which is necessary to exclude unrelated fluctuations. Techniques built for non-stationary signals \citep[e.g.,][]{2022SSRv..218....9A} can overcome these difficulties. One of them, Empirical Mode Decomposition \citep[EMD;][]{Huang1998}, has been used on QPPs by, for example, \citet{Kolotkov2015}, \citet{2019ApJS..244...44B} and \citet{2026ApJ..1003...78K}, and \citet{2016ApJ...830..110C} applied it to demonstrate that, for pulsations in the decay phase of solar and stellar soft X-ray flares, the damping time is proportional to the period.
	
	The hard X-ray spectrometer HEL1OS \citep{Nandi.Aditya2025} on board Aditya-L1 \citep{Parate.Aditya2025} observes flares continuously in several energy bands at a rapid cadence, so it is well matched to this problem. \citet{2026arXiv260922787R} searched 34 X-class flares seen by HEL1OS between July 2024 and March 2026 and found significant pulsations in 25 of them. In this work we build on that catalogue by comparing the energy dependence of the modulation depth of QPPs in M- and X-class flares. The M-class sample comprises 18 events, and the X-class sample is taken from \citet{2026arXiv260922787R}. The modulation depth is measured in four HEL1OS channels covering 8--20~keV. The number of usable measurements differs from channel to channel. Our goals are (i) to determine how the mean and median modulation depth changes with photon energy in each flare class, (ii) to establish whether the energy dependence differs systematically between M- and X-class flares, and (iii) to assess how far any such difference can be trusted given the unequal channel-specific sample sizes and the large event-to-event scatter. The comparison is expected to indicate whether the amplitude of QPPs depends on flare class, which in turn helps to constrain whether the pulsations are mainly produced by MHD modulation of the emission or by self-sustained, repetitive reconnection, and it lays the groundwork for more rigorous statistical tests based on a uniform event selection and consistent uncertainty estimates.\\
	
	The remainder of the paper is organised as follows. The instruments and data are described in Section~\ref{sec-obs-data}. The results and their interpretation are given in Section~\ref{sec-results-discussions}, and the main conclusions in Section~\ref{sec-summary}.

	\section{The Observational Data}
	\label{sec-obs-data}

	The hard X-ray measurements analysed here were made by the High Energy L1 Orbiting X-ray Spectrometer (HEL1OS), one of the payloads of the Aditya-L1 observatory stationed at the first Sun--Earth Lagrange point \citep{Nandi.Aditya2025, Parate.Aditya2025}. HEL1OS is dedicated to solar flare X-rays and spans a broad band from 8 to 150~keV. To cover it, the instrument combines two types of semiconductor sensor with different strengths. Cadmium telluride (CdTe) detectors respond to the lower part of the band, from 8 up to 70~keV, while pixelated cadmium zinc telluride (CZT) modules are sensitive from 20 to 150~keV and are better suited to harder photons.
	
	The energy scale is calibrated in flight with a radioactive $^{241}\mathrm{Am}$ source, using the well-defined 59.5~keV line that it emits. At this level of calibration, the CdTe sensors resolve spectral structure to about 1~keV at low energies (around 14~keV), whereas the CZT modules are coarser, at roughly 7~keV near 60~keV. Photon arrival times can in principle be recorded to 10~ms. However, the routinely released science products are accumulated over 1~s intervals, and this is the time resolution used in our analysis.


	\section{Results and Discussions}
	\label{sec-results-discussions}

	\begin{figure*}[b]
		\centering
		\includegraphics[width=0.45\textwidth,clip]{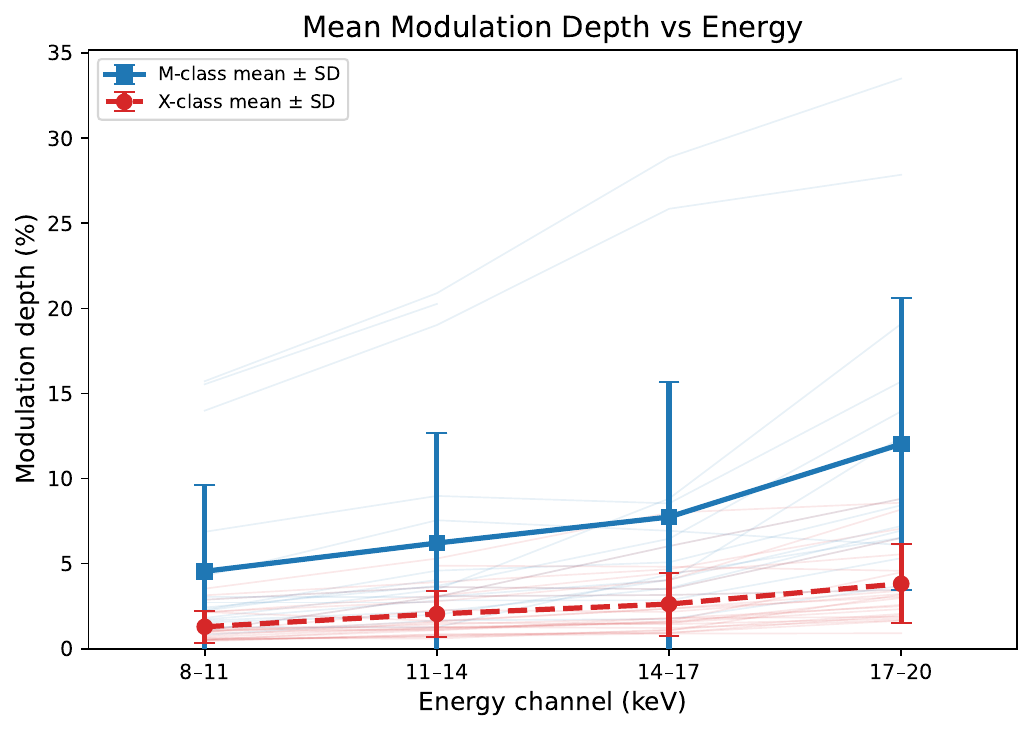}
		\includegraphics[width=0.45\textwidth,clip]{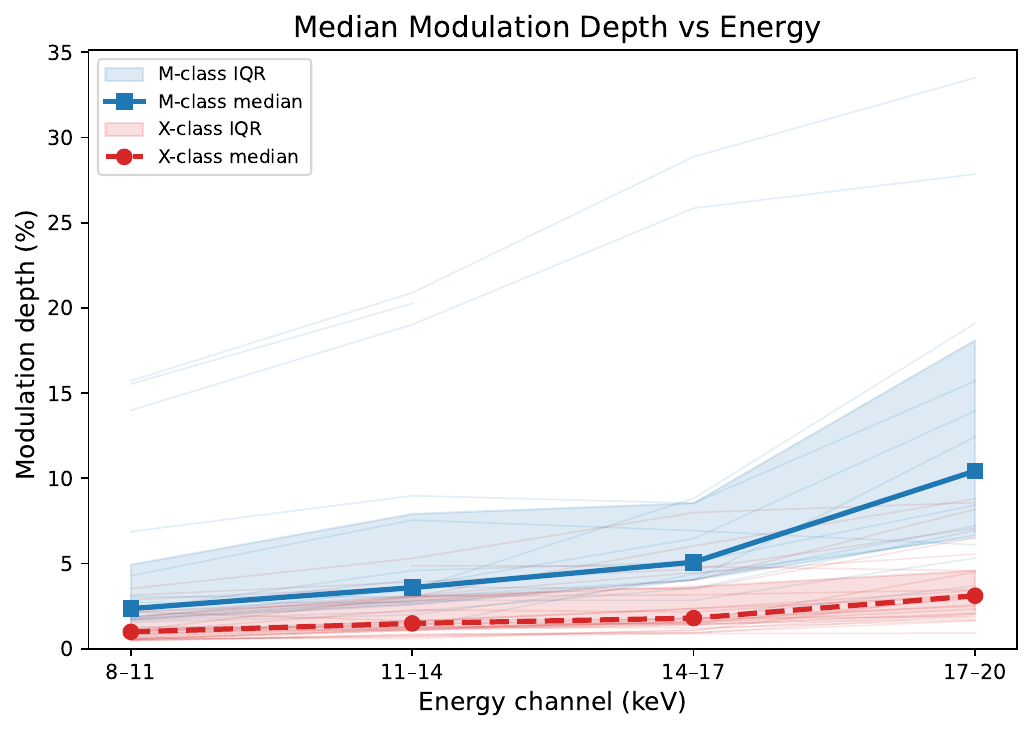}
		\caption{Energy dependence of the modulation depth of hard X-ray quasi-periodic pulsations (QPPs) in M- and X-class solar flares observed by HEL1OS. The left panel presents the mean modulation depth in each energy channel, with error bars indicating the standard deviation (SD); the right panel shows the corresponding median values and interquartile ranges (IQRs). Blue solid curves and square markers represent the M-class sample, whereas red dashed curves and circular markers represent the X-class sample. Faint lines show the modulation-depth measurements for individual flares. The annotations indicate the number of events contributing to each energy-channel distribution. The two statistical representations provide complementary views of the central tendency and dispersion of the measurements, while the energy-dependent sample sizes should be considered when interpreting differences between flare classes.}
		\label{fig:md_mx_mean_median}
	\end{figure*}

	To investigate how the modulation depth of quasi-periodic pulsations (QPPs) varies with flare class and photon energy, we compare the modulation depth (MD) measured in M-class flares with the corresponding measurements for X-class flares. The X-class results are taken from our previous study \citep{2026arXiv260922787R}, while the M-class measurements are obtained from the HEL1OS hard X-ray light curves in four energy bands: 8–11, 11–14, 14–17, and 17–20 keV.
	
	The QPP signatures are identified using ensemble empirical mode decomposition (EEMD), which separates a non-stationary light curve into intrinsic mode functions (IMFs), followed by wavelet analysis to investigate the time-dependent periodic behaviour. The dominant periodicity is determined from the global wavelet spectrum \citep{Torrence1998} of the selected IMF, with its statistical significance assessed against a red-noise background. The modulation depth is then calculated over the corresponding QPP interval using the same approach adopted in our previous study \citep{2026arXiv260922787R}. The event-wise periodicities, their uncertainties, and the associated modulation depths for the M-class sample are listed in Table~\ref{tab:events-list-qpp-m-class}.
	
	Figure~\ref{fig:md_mx_mean_median} compares the modulation-depth distributions of the M- and X-class samples as a function of photon energy. The left panel presents the mean modulation depth with its standard deviation, whereas the right panel shows the median and interquartile range. The faint curves represent individual-event measurements, allowing the event-to-event variability to be examined alongside the overall trends. The numbers annotated in the figure indicate the number of events contributing to each energy band.
	
	The mean and median distributions indicate that the modulation depth generally increases towards higher photon energies in both flare classes. However, the M-class sample exhibits a broader spread and, in several energy bands, higher central modulation-depth values than the X-class sample. 
	
	The mean and median provide complementary descriptions of the samples. The mean is sensitive to events with comparatively large modulation depths, whereas the median is less affected by extreme values. Consequently, agreement between the two measures provides a useful indication of whether the overall trend is representative of the central distribution or is strongly influenced by a small number of events. The individual-event measurements in Table~\ref{tab:events-list-qpp-m-class} also demonstrate why the spread of the distributions should be considered when comparing the two flare classes.
	
	These trends should not, by themselves, be interpreted as evidence that M-class flares intrinsically produce stronger QPPs than X-class flares. Modulation depth measures the relative amplitude of a periodic component rather than its absolute hard X-ray power. A physical interpretation therefore requires consideration of the background emission, the duration and significance of the detected oscillations, the energy dependence of the thermal and non-thermal contributions, and the event-selection criteria. In addition, the number of usable events varies across energy bands, which should be taken into account when assessing the robustness of the comparison.
	
	Overall, the comparison provides an observational basis for examining the relationship between flare class, photon energy, and the relative amplitude of hard X-ray QPPs.

	\section{Summary and Conclusions}
	\label{sec-summary}

	We have investigated the energy dependence of the modulation depth of hard X-ray (HXR) quasi-periodic pulsations (QPPs) in M- and X-class solar flares observed by HEL1OS onboard Aditya-L1. Using measurements in four energy channels spanning 8--20~keV, we compared the distributions of QPP modulation depth between the two flare populations. The analysis extends our previous investigation of X-class flares \citep{2026arXiv260922787R} by examining how the relative amplitude of HXR QPPs varies with photon energy in M-class flares and by comparing the distributions across the two flare classes. The event-wise M-class measurements, including the detected periodicities, their uncertainties, and modulation depths, are provided in Table~\ref{tab:events-list-qpp-m-class}.
	
	The principal findings of this study are summarised as follows:
	
	\textit{Energy-dependent modulation depth:}
	The median modulation depth increases with photon energy in both flare populations. For the M-class sample, the mean value rises from approximately 2.3\% in the 8--11~keV channel to approximately 13.2\% in the 17--20~keV channel. The corresponding mean for the X-class sample increase from approximately 1.0\% to approximately 3.1\%. Thus, both populations exhibit larger relative QPP modulation at higher photon energies, consistent with previous HXR and multi-wavelength investigations reporting enhanced pulsation amplitudes at higher energies \citep{Li2008, NakariakovMelnikov2009, Kupriyanova2010, Hayes2016, Li2017}. Reviews of solar QPPs have also discussed energy-dependent oscillation amplitudes in the context of periodic energy release and electron acceleration \citep{Kupriyanova2020, McLaughlin2018}.
	
	\textit{Contrast between M- and X-class flares:}
	The M-class sample exhibits a more pronounced increase in modulation depth towards the highest energy channel than the X-class sample. This is the main observational contrast identified in our comparison. However, modulation depth quantifies the relative variability of the emission rather than its absolute oscillation amplitude or total HXR output. The observed contrast therefore does not, on its own, establish that M-class flares produce intrinsically stronger QPPs. Differences in background emission, event-to-event variability, and the number of usable measurements in each energy channel may influence the comparison.

	In conclusion, the HEL1OS observations reveal an increasing median HXR QPP modulation depth with photon energy in both M- and X-class flares, accompanied by a more pronounced rise towards 17--20~keV in the M-class sample.

	\section{Acknowledgment}
	
	Aditya-L1 is an observatory class mission which is fully funded and operated by the Indian Space Research Organization (ISRO). The mission was conceived and realised with the help from various ISRO centres. The science payloads and science ready data products are realised by the payload PI institutes in close collaboration with ISRO centres. The PI institutes are: Indian Institute of Astrophysics (IIA); Inter University Centre for Astronomy and Astrophysics (IUCAA), Laboratory for Electro-optics Systems (LEOS/URSC); Physical Research Laboratory (PRL); U R Rao Satellite Centre (URSC); and Space Physics Laboratory (SPL/VSSC). We acknowledge the use of data from the Aditya-L1 mission of the Indian Space Research Organisation (ISRO), archived at the Indian Space Science Data Centre (ISSDC). One of the authors, R.~A.~M., gratefully acknowledges support from the ISRO RESPOND project (No. ISRO/RES/2/437/21-22). V.M.N. is supported by ERC grant 101201424 (ACDCSUN) and the BK21 FOUR program through the National Research Foundation of Korea (NRF) under the Ministry of Education (MoE) (Kyung Hee University, Human Education Team for the Next Generation of Space Exploration), and the Global-Learning \& Academic research institution for Master’s/PhD students, the Postdocs (G-LAMP) Program of the NRF grant funded by the MoE (RS-2025-25442355). 
	
	\appendix

	\small
\setlength{\tabcolsep}{4pt}

\begin{longtable}{|c|c|c|c|cc|cc|}
	\caption{Catalog of QPP parameters derived from HEL1OS observations of M-class solar flares. P1 and P2 denote the first and second detected periodicities (period $\pm$ uncertainty) with their modulation depths (MD).}
	\label{tab:events-list-qpp-m-class}\\
	
	\hline
	\multicolumn{3}{|c|}{\textbf{Event}} &
	\multicolumn{5}{c|}{\textbf{HEL1OS}}\\
	\hline
	S.N. &
	Date &
	Flare &
	Energy &
	\multicolumn{2}{c|}{P1} &
	\multicolumn{2}{c|}{P2}\\
	
	\cline{5-8}
	
	&&&
	(keV) &
	Period (min) & MD (\%) &
	Period (min) & MD (\%)\\
	\hline
	\endfirsthead
	
	\multicolumn{8}{c}{\tablename\ \thetable{} -- continued from previous page}\\
	\hline
	\multicolumn{3}{|c|}{\textbf{Event}} &
	\multicolumn{5}{c|}{\textbf{HEL1OS}}\\
	\hline
	S.N. &
	Date &
	Flare &
	Energy &
	\multicolumn{2}{c|}{P1} &
	\multicolumn{2}{c|}{P2}\\
	
	\cline{5-8}
	
	&&&
	(keV) &
	Period (min) & MD (\%) &
	Period (min) & MD (\%)\\
	\hline
	\endhead
	
	\hline
	\multicolumn{8}{r}{Continued on next page}\\
	\hline
	\endfoot
	
	\hline
	\endlastfoot

	1 & 17 Jul 24 & M5.0 &
	8--11 &
	$1.30 \pm 0.27$ & 2.72 &
	&\\

	& & &
	11--14 &
	$1.16 \pm 0.20$ & 2.72 &
	&\\
	\hline

	2 & 24 Jul 24 & M2.9 &
	8--11 &
	$1.23 \pm 0.29$ & 1.81 &
	&\\

	& & &
	11--14 &
	$1.38 \pm 1.19$ & 3.48 &
	&\\

	& & &
	14--17 &
	$1.74 \pm 0.50$ & 8.82 &
	&\\

	& & &
	17--20 &
	$1.95 \pm 0.48$ & 19.09 &
	&\\
	\hline

	3 & 28 Jul 24 & M7.8 &
	8--11 &
	$1.95 \pm 0.54$ & 2.42 &
	&\\

	& & &
	11--14 &
	$1.95 \pm 0.39$ & 3.69 &
	&\\

	& & &
	14--17 &
	$2.06 \pm 0.45$ & 6.46 &
	&\\

	& & &
	17--20 &
	$2.06 \pm 0.48$ & 13.95 &
	&\\
	\hline

	4 & 29 Jul 24 & M8.7 &
	8--11 &
	$1.64 \pm 0.49$ & 1.55 &
	&\\

	& & &
	11--14 &
	$1.74 \pm 0.52$ & 1.99 &
	&\\

	& & &
	14--17 &
	$1.74 \pm 0.30$ & 4.05 &
	&\\

	& & &
	17--20 &
	$1.95 \pm 0.86$ & 7.21 &
	&\\
	\hline

	5 & 31 Jul 24 & M5.4 &
	8--11 &
	$1.38 \pm 0.46$ & 1.65 &
	$2.92 \pm 1.34$ & 1.26\\

	& & &
	11--14 &
	$1.38 \pm 0.58$ & 2.19 &
	$3.09 \pm 1.02$ & 2.46\\

	& & &
	14--17 &
	$1.38 \pm 0.29$ & 3.56 &
	$2.75 \pm 0.48$ & 3.10\\

	& & &
	17--20 &
	$1.95 \pm 0.58$ & 6.94 &
	$3.09 \pm 0.81$ & 5.41\\
	\hline

	6 & 1 Aug 24 & M8.2 &
	8--11 &
	$1.23 \pm 0.39$ & 1.72 &
	$3.90 \pm 0.99$ & 1.54\\

	& & &
	11--14 &
	$1.55 \pm 0.51$ & 2.84 &
	$4.13 \pm 1.53$ & 1.91\\

	& & &
	14--17 &
	$1.23 \pm 0.35$ & 2.84 &
	$2.75 \pm 1.78$ & 2.35\\

	& & &
	17--20 &
	$1.74 \pm 0.49$ & 5.32 &
	$2.45 \pm 1.03$ & 3.49\\
	\hline

	7 & 5 Aug 24 & M6.1 &
	17--20 &
	$2.60 \pm 1.12$ & 18.90 &
	&\\
	\hline

	8 & 14 Oct 24 & M3.4 &
	8--11 &
	$1.09 \pm 0.54$ & 2.24 &
	&\\

	& & &
	11--14 &
	$1.38 \pm 0.45$ & 4.04 &
	&\\
	\hline

	9 & 15 Oct 24 & M2.1 &
	8--11 &
	$1.84 \pm 0.73$ & 6.87 &
	&\\

	& & &
	11--14 &
	$1.84 \pm 0.82$ & 8.98 &
	&\\

	& & &
	14--17 &
	$1.64 \pm 0.29$ & 8.54 &
	&\\

	& & &
	17--20 &
	$1.74 \pm 0.57$ & 15.70 &
	&\\
	\hline

	10 & 17 Oct 24 & M2.4 &
	8--11 &
	$1.55 \pm 0.51$ & 3.05 &
	&\\

	& & &
	11--14 &
	$1.30 \pm 0.77$ & 3.04 &
	&\\

	& & &
	14--17 &
	$1.84 \pm 0.32$ & 4.04 &
	&\\

	& & &
	17--20 &
	$1.74 \pm 0.40$ & 12.43 &
	&\\
	\hline

	11 & 18 Oct 24 & M4.7 &
	No QPP &
	&
	&
	&
	\\
	\hline

	12 & 13 Dec 24 & M1.0 &
	8--11 &
	$1.84 \pm 0.43$ & 15.54 &
	&\\

	& & &
	11--14 &
	$1.95 \pm 0.53$ & 20.26 &
	&\\
	\hline

	13 & 24 Dec 24 & M4.1 &
	8--11 &
	$1.74 \pm 0.43$ & 15.72 &
	&\\

	& & &
	11--14 &
	$1.74 \pm 0.43$ & 20.89 &
	&\\

	& & &
	14--17 &
	$1.74 \pm 0.43$ & 28.87 &
	&\\

	& & &
	17--20 &
	$1.74 \pm 0.36$ & 33.50 &
	&\\
	\hline

	14 & 3 Feb 25 & M3.1 &
	8--11 &
	$1.95 \pm 0.54$ & 13.99 &
	$4.37 \pm 1.08$ & 13.06\\

	& & &
	11--14 &
	$1.95 \pm 0.54$ & 19.02 &
	$4.37 \pm 1.18$ & 15.17\\

	& & &
	14--17 &
	$1.95 \pm 0.51$ & 25.85 &
	$4.13 \pm 1.12$ & 12.53\\

	& & &
	17--20 &
	$1.95 \pm 0.51$ & 27.85 &
	$3.90 \pm 1.28$ & 9.82\\
	\hline

	15 & 23 Feb 25 & M4.9 &
	8--11 &
	$1.64 \pm 0.54$ & 2.29 &
	&\\

	& & &
	11--14 &
	$1.74 \pm 0.82$ & 4.59 &
	&\\

	& & &
	14--17 &
	$1.74 \pm 0.72$ & 5.08 &
	&\\

	& & &
	17--20 &
	$1.38 \pm 0.24$ & 8.43 &
	&\\
	\hline

	16 & 25 May 25 & M8.9 &
	8--11 &
	$2.60 \pm 1.38$ & 4.28 &
	&\\

	& & &
	11--14 &
	$2.92 \pm 1.15$ & 7.55 &
	&\\

	& & &
	14--17 &
	$2.92 \pm 0.97$ & 6.93 &
	&\\

	& & &
	17--20 &
	$2.92 \pm 0.85$ & 6.10 &
	&\\
	\hline

	17 & 15 Jun 25 & M8.4 &
	8--11 &
	$1.30 \pm 0.29$ & 0.82 &
	$4.63 \pm 1.66$ & 1.06\\

	& & &
	11--14 &
	$1.16 \pm 0.51$ & 1.67 &
	$3.90 \pm 1.37$ & 1.71\\

	& & &
	14--17 &
	$1.23 \pm 0.26$ & 1.74 &
	$3.68 \pm 0.60$ & 2.13\\

	& & &
	17--20 &
	$1.74 \pm 0.32$ & 3.70 &
	$2.75 \pm 1.03$ & 3.70\\
	\hline

	18 & 3 Nov 25 & M5.0 &
	8--11 &
	$1.38 \pm 0.37$ & 1.28 &
	$3.68 \pm 0.64$ & 1.06\\

	& & &
	11--14 &
	$1.38 \pm 0.60$ & 1.26 &
	$3.47 \pm 0.60$ & 2.51\\

	& & &
	14--17 &
	$2.06 \pm 0.68$ & 4.37 &
	$3.28 \pm 0.57$ & 4.49\\

	& & &
	17--20 &
	$2.19 \pm 0.51$ & 6.53 &
	$3.68 \pm 1.14$ & 14.30\\
	\hline

\end{longtable}


\end{document}